# An interactive framework for the evaluation and detection of stereoacuity threshold under ambient lighting

Kritika Lohia[1], Rijul S. Soans[1,2], Rohit Saxena[3], Tapan K. Gandhi[1]

[1]Department of Electrical Engineering, Indian Institute of Technology – Delhi, New Delhi, India

[2]Herbert Wertheim School of Optometry and Vision Science, University of California, Berkeley, USA.

[3]Dr. Rajendra Prasad Centre for Ophthalmic Sciences, All India Institute of Medical Sciences, New Delhi, India

**Support:** KL was supported by the Indian Institute of Technology - Delhi, India. RSS was supported by the Visvesvaraya PhD Scheme, MEITY, Govt. of India: MEITY-PHD-1810. RS and TKG have received funding from the Department of Science and Technology - Cognitive Science Research Initiative Project #RP03962G, Govt. of India. TKG was also supported by the Cadence Chair for Automation & AI, RP04329G, and SERB-INAE RP04613G Govt. of India. The funding organizations had no role in the design, conduct or analysis of this research.

**Additional Information:** Author RSS was affiliated with IIT-Delhi when the majority of the work described in this manuscript was done. He is currently affiliated with UC Berkeley.

**Conflict of Interest:** KL, RSS, RS and TKG are listed as inventors on the Indian complete patent application #202411045008 "T.K. Gandhi, K. Lohia, R.S. Soans, R. Saxena (2024), "Method and system for evaluating a stereoacuity threshold of a user in ambient lighting", which is based on the methods described in this manuscript.

## Abstract

**Objective:** Our study aims to provide a novel framework for the continuous evaluation of stereoacuity under ambient lighting conditions using Bayesian inference.

**Methods:** We applied a combination of psychophysical and expected entropy minimization procedures for the computation of a continuous stereoacuity threshold. Subsequently, we evaluated the effect of ambient lighting during stereoacuity testing (ST) by adopting a bisection-matching based adaptive gamma calibration (AGC). Participants (N=187) including visually healthy controls (N=51), patients with Intermittent Divergent Squint (IDS; N=45), and controls with induced anisometropia (IA; N=91) performed ST with and without AGC under two lighting conditions: completely dark (20 cd/m$^2$) and normally lit (130 cd/m$^2$) rooms.

**Results:** Our framework demonstrated "excellent" reliability (> 0.9) and a positive correlation with TNO (a clinical stereo test), regardless of whether AGC was conducted. However, when AGC is not performed, significant differences (Friedman $X_r^2$ = 28.015; p<0.00001; Bland-Altman bias: 30 arc-sec) were found in stereoacuity thresholds between dark and light conditions for participants with IDS and IA. Controls are unaffected by AGC and yield a similar stereoacuity threshold under both lighting conditions.

**Conclusion:** Our study proves that stereoacuity threshold is significantly deviated particularly in participants with IDS or IA stereo-deficits if ambient lighting is not taken into consideration. Moreover, our framework provides a quick (approximately 5-10 minutes) assessment of stereoacuity threshold and can be performed within 30 ST and 15 AGC trials.

**Significance:** Our test is useful in planning treatments and monitoring prognosis for patients with stereo-deficits by accurately assessing stereovision.

## 1. Introduction

Stereopsis is an integral factor in the clinical evaluation of a patient's vision. The need for both eyes to work together to extract information from the surroundings is crucial in performing day-to-day activities such as reading, navigating, judging distances, and eye-hand coordination[1–4]. Consequently, stereopsis is considered the benchmark of the peak performance of clinical vision and is the highest marker of quality of life[5–7]. Moreover, evaluating stereopsis provides valuable insights into the integrity of the visual pathway and binocular function. Impairments in stereopsis can indicate underlying disorders such as strabismus[8] and amblyopia[9], which might otherwise remain undetected.

Several tests have been developed to assess stereopsis - all sharing the core principle of presenting distinct images to each eye but with a slight distinction in the way the disparity is presented. For example, the TNO test uses color filter separation, the RANDOT test uses polarized light, the Lang test uses lenticular lenses to create the disparity between the images, and the Frisby-Davis test uses physical objects arranged at different distances. These stereoacuity assessment tests are either available as printed books or charts (except for the Frisby Davis test, which is a real-depth test) in clinical settings. Although clinically useful, these standard tests have some limitations. Firstly, while presentation methods, in principle, should not affect disparity detection, there is evidence that the normative data varies across these tests - implying that the way disparity is presented can impact the results[10]. Moreover, such variations are not only limited to the type of stereo test used but also within the same test - for example, in the case of TNO - where it was reported that the stereoacuity outcomes differed with manufacturer changes in the printed books[11]. Secondly, these tests measure stereoacuity at discrete levels (for example, at 480, 240, 120 and, 60 arc-secs in TNO). This leads to the inaccurate measurement of stereopsis, particularly during the post-operative follow-up in strabismus and in the prognosis of other refractive error-based vision conditions[12]. Lastly, these printed book tests are expensive, difficult to procure in low-resource ophthalmic settings and involve recurrent costs because their pages fade over time.

Computerized versions of the traditional stereo tests have addressed the above limitations to some extent - particularly those involving Random Dot Stereograms - ASTEROID v0.9[13] & v1[14], eRDS v6[15], Digital test[16] and others[17–19]. However, these tests are limited in two aspects – 1) none of them have incorporated the effects of ambient lighting conditions on stereopsis. It has been well documented that ambient lighting levels can significantly affect the perceived contrast when viewed digitally[20–23], and therefore can lead to false judgment and reliability of stereoacuity thresholds[24–26]. 2) It is difficult to truly judge the clinical validity of these computerized tests because of the difference in measurement scales i.e. the clinical tests provide discrete stereoacuity values, whereas these tests are in a continuous measurement scale.

While the stereoacuity thresholds provided by ASTEROID v1 closely align with those of the Randot Circles test, it is important to note that the two tests fundamentally differ in nature. ASTEROID v1 does not have monocular cues, whereas the Randot Circles test includes monocular contours[27]. Therefore, in the context of clinical validity, a true comparison cannot be made. Although the eRDS v6 test provides better precision, this test is burdened with long measurement times (45 minutes) and needs a mirror-based stereoscope. Furthermore, both ASTEROID and eRDS v6 are tested only on visually healthy controls, which might lead to a false perception of agreement with the clinical stereotests. In fact, Tittes et al.[16] found variability in the agreement levels when they tested their computer-based stereoacuity test and the TNO on populations with and without stereo-deficits. Thus, it is imperative that the procedures used for evaluating stereopsis are meticulously conducted and adhere to proper methodology.

Therefore, we propose a new framework that – 1) addresses the limitations of the clinical stereotests, specifically those relying on expensive printed book/chart technology. The limitations, such as – variations in different editions of printed books, can often result in clinically unacceptable differences in measurements and challenges in terms of procurement and the associated recurrent costs, 2) is fast and efficient in providing the stereoacuity threshold on a continuous measurement scale using Bayesian inference. 3) is reliable across a wide range of target groups, including visually healthy and stereoscopically impaired population, 4) incorporates ambient lighting using Adaptive Gamma Calibration (AGC) for the evaluation of stereoacuity.  To validate our framework, we recruited participants to perform Stereoacuity Testing (ST) with and without adjusting for ambient lighting and under two lighting conditions: completely dark (20 $cd/m^2$) and normally lit (130 $cd/m^2$) rooms. This gives rise to two modes of testing – ST only (which we refer as 1-step paradigm) and performing AGC before ST (which we refer as 2-step paradigm) under two lighting conditions – dark and light. We hypothesized that ambient light affects stereoacuity if there is a significant difference in stereoacuity thresholds measured across dark and light modes in the 1-step paradigm, and there is no significant difference in stereoacuity thresholds measured across dark and light modes in the 2-step paradigm. If our hypothesis is proven to be true, then our framework would essentially provide a novel, interactive and accurate measure for the detection and evaluation of stereoacuity thresholds under ambient lighting.

## 2. Materials and Methods

Table 1 describes the characteristics of the all participants included in this study. For the 2-step paradigm (AGC and ST), we recruited 27 visually healthy controls, 23 patients with Intermittent Divergent Squint (IDS) and 47 visually healthy controls with Induced Anisometropia (IA). For the 1-step paradigm (ST alone), we recruited another set of 24 visually healthy controls, 22 patients with IDS and 44 visually healthy controls with IA. Anisometropia was induced by adding trial lenses from -4.75D to +0.75D spherical and from -1 to +1D cylindrical lenses to participant’s baseline refractive correction in a trial frame over the non-dominating eye to reduce their stereoacuity. While the purpose of including IDS and IA was to expand the range of stereoacuity values, therefore, we did not differentiate between the two categories for all statistical analyses. Instead, we compared the performance of participants (1-step: N=66; 2-step: N=70) with real (IDS) or induced stereo-deficits (IA) against the visually healthy controls. If a participant's stereoacuity after inducing anisometropia exceeded the TNO measurement scale (poorer than 480 arc-sec), it was recorded as 1663 arc-sec (i.e., the maximum value measured by ST), regardless of whether AGC was performed. All recruited participants were required to have a Best Corrected Visual Acuity (BCVA) of 6/9 (0.67 or $\leq$ 0.17 logMAR) or better in both eyes. All participants provided their written or verbal consent to participate in this study, and the ethics committee of All India Institute of Medical Sciences-Delhi and Indian Institute of Technology-Delhi, New Delhi, India. This study is in accordance with the tenets of the Declaration of Helsinki.

| Population | N | Type | No. of Males | Median Stereoacuity (arc-sec) | Mean age (years) |
|---|---|---|---|---|---|
| Controls | 27 | 2-step | 16 | 60 | 27.92±3.6 |
| | 24 | 1-step | 16 | 60 | 24.95±3.92 |
| Controls with IA | 47 | 2-step | 27 | 480 | 27.34±4.77 |
| | 44 | 1-step | 28 | 480 | 27.7±2.54 |
| IDS | 23 | 2-step | 13 | 60 | 28.08±5.78 |
| | 22 | 1-step | 13 | 60 | 28.95±6.09 |

Table 1. Characteristics of the participants included in this study.

### 2.1 Stimulus Apparatus

We utilized 3D anaglyph glasses with red film over the left eye and cyan film over the right eye. We displayed our stimuli on Dell XPS 15 7590 with a 15-inch display having a native resolution of 3840 × 2160, a refresh rate of 60 Hz and a dot pitch of 0.0864. However, the stimuli were presented at a display resolution of 800 × 600 pixels resulting in an effective horizontal screen dimension of 25.8 cm. The scripts used to create the texture scenes and the stimuli renderings were conducted using MATLAB R2020b and Psychtoolbox v.3.0.17.

### 2.2 Creation of the Adaptive Gamma Calibration Stimuli

Here, we describe the creation of AGC stimuli which will be utilized in the 2-step paradigm. The process for the adjustment of ambient lighting using AGC is described in the next section. The AGC stimuli consisted of three adjacent textures, namely – left, middle and right, on a white background (see Figure 1). The left and right textures (hereafter collectively referred to as mixed intensity textures) consisted of 128 alternating lines of known high (bright) and low (dark) intensity values (as described in Equations 2 to 7). The number of alternating high and low-intensity lines can be customized to accommodate different viewing distances, texture size and display size (width and resolution). The width of each line is 1 pixel. The middle texture (also called the calibration texture) is filled with a single gray intensity value and subtends an angle of 5.72° at a distance of 40 cm. The procedure to compute the gray intensity value of the middle texture is described in the subsequent section.

### 2.3 Adaptive Gamma Calibration Procedure

For the AGC task, we used the method of bisection[28,29]. The participants had to determine the point of subjective equality between the calibration texture and the mixed-intensity textures. Specifically, the participants had to adjust the gray intensity value of the middle texture such that the perceived luminance of the middle texture matched the luminance of the mixed-intensity textures in a total of 15 trials. The intensity adjustment of the middle texture could be made in two modes – coarse and fine steps, wherein each step is of value 3/255 and 0.3/255, respectively. For each trial, the coarse and fine adjustments were done using the left or right and up or down arrow keys, respectively.  For example, if the up key is pressed, the intensity is increased by an amount of 0.3/255. Similarly, if the left key is pressed the intensity is decreased by an amount of 3/255. Assuming that the participant is taking at most 8-12 seconds per trial, the time taken to complete the entire AGC procedure is approximately 2-3 minutes. An example of one step adjustment is shown in Figure 1 (when compared to Figure 2).

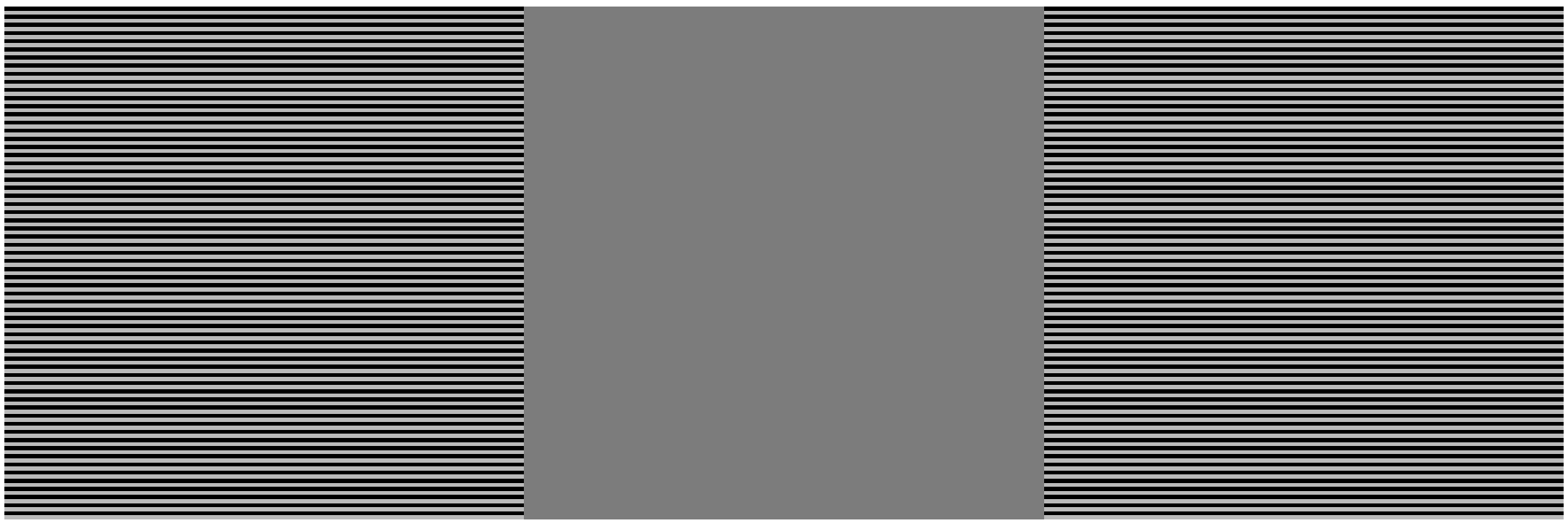

Figure 1. An example of the AGC stimuli trial having three adjacent textures namely – left, middle and right on a white background.

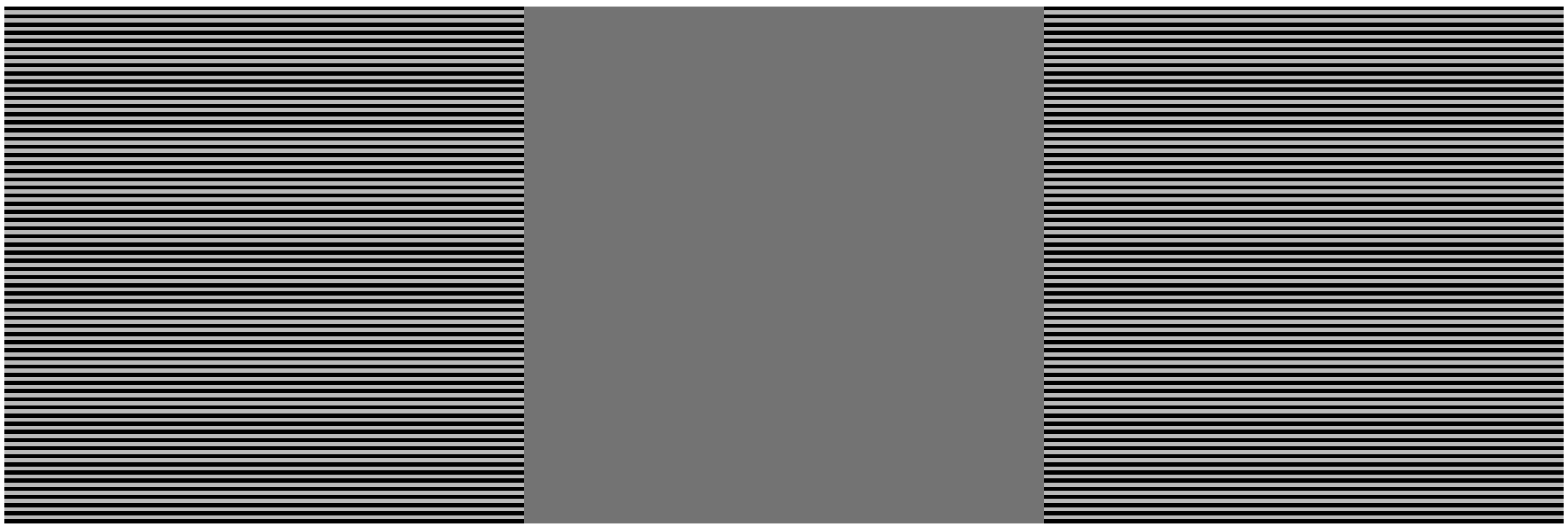

Figure 2. An example of one step adjustment of the middle texture when compared to Figure 1.

Below, we describe the procedure to compute gamma in detail. The general relationship between gray values and the actual luminance (as measured with photometer) is described as power-law transformation:

$$I = Lum^{(\frac{1}{\gamma_0})} \qquad (1)$$

Wherein,

$I$ represents gray level values, $Lum$ represents the actual luminance values and $\gamma_0$ is the initial gamma of the display (which is taken as 2.2 as a standard gamma on the Windows OS).

Since AGC had to be performed without a photometer, we selected *N* preset values for luminance. For demonstration, we chose a luminance vector having *N=17* evenly distributed values between 0 and 1 (shown below in Equation 2); however, *N* is customizable.

$Lum =$ {0 0.0625 0.125 0.1875 0.25 0.3125 0.375 0.4375 0.5 0.5625 0.625 0.6875 0.75 0.8125 0.875 0.9375 1} (2)

Next, we computed the corresponding gray level values $I$ by substituting Equation 2 into Equation 1.

$I$ = {0 0.283 0.388 0.467 0.532 0.589 0.640 0.686 0.729 0.769 0.807 0.843 0.877 0.909 0.941 0.971 1} (3)

Equations 2 and 3 can be collectively seen as a lookup table. Further, to compute the intensity values in mixed intensity textures, we predefined two vectors – high and low index – spanning all trials. These are given below:

$$h_v = \{17\ 17\ 9\ 17\ 13\ 9\ 5\ 17\ 15\ 13\ 11\ 9\ 7\ 5\ 3\} \qquad (4)$$

$$l_v = \{1\ 9\ 1\ 13\ 9\ 5\ 17\ 15\ 13\ 11\ 9\ 7\ 5\ 3\ 1\} \qquad (5)$$

where $h_v$ and $l_v$ are the high and low index vectors, respectively.

The high and low intensity values at trial *i* are then computed as:

$$I_{high}(i) = I(h_v\ (i)) \qquad (6)$$

$$I_{low}(i) = I(l_v\ (i)) \qquad (7)$$

At each trial, the middle texture is initially displayed as an average of the mixed intensity texture given by Equation 8.

$$I_{init}(i) = \left(\frac{I_{high}(i)+I_{low}(i)}{2}\right) \qquad (8)$$

If the perceived luminance of the middle texture is not same as that of the mixed textures, then the participant performs coarse or fine adjustments and a new intensity $(I_{new})$ is assigned for the middle texture. Otherwise, the participant directly proceeds to the next trial by pressing the spacebar key.

$$I_{new}(i) = I_{init}(i) \pm step(i) \qquad (9)$$

The error between the actual intensity of the calibration texture and the intensity perceived by the participant is given by the difference between Equations 8 and 9.

$$Error\ (i) = I_{init}(i) - I_{new}(i) \qquad (10)$$

A new gamma $\gamma_{new}$ is computed by minimizing the sum-of-squared-errors across all trials.

$$\gamma_{new} = \min \sum_{i=1}^{15} (I_{init}(i) - I_{new}(i))^2 \qquad (11)$$

Next, we compute the normalized luminance vector by:

$$Lum_{norm} = G^{\gamma_{new}} \qquad (12)$$

Where,

$G$ is the vector with values ranging from 0-255, representing 8-bit grayscale.

Finally, the normalized luminance values are updated in the display parameters using *Screen('LoadNormalizedGammaTable')* command of the Psychtoolbox v.3.0.17. After performing this step, we have effectively adjusted the gamma of the display according to available ambient lighting for a particular user.

**2.4 Creation of the Random dot stereogram stimuli**

The RDS stimuli consisted of two overlapping red and cyan colored textures each of size 8.6 cm on a black background. These overlapping textures subtended 12.27° when viewed at a distance of 40 cm. The size and viewing distance were chosen in accordance with the clinical stereotest TNO. Each overlapping texture contained 30000 random dots wherein the size of each dot was 0.0089 pixels. Inside each overlapping texture, there was a hidden texture of the shapes – '⊔', '⊓', '⊏' and '⊐'. The hidden texture consisted of 8400 random dots. The pixel size of the red, cyan and the hidden textures can be customized to incorporate different viewing distances, display resolution and physical dimensions. However, the size of red and cyan textures have to be maintained at the same level with respect to each other. The hidden textures were horizontally shifted with an effective offset $(O_1)$ ranging from 0.1 pixel to 10 pixels amounting to 16.63 to 1663 arc-sec when viewed at 40 cm. We used anti-aliasing to represent sub-pixel level of shift in pixels. Figure 3 illustrates the geometry of the perceived protruding shape ‘⊔’as an example.

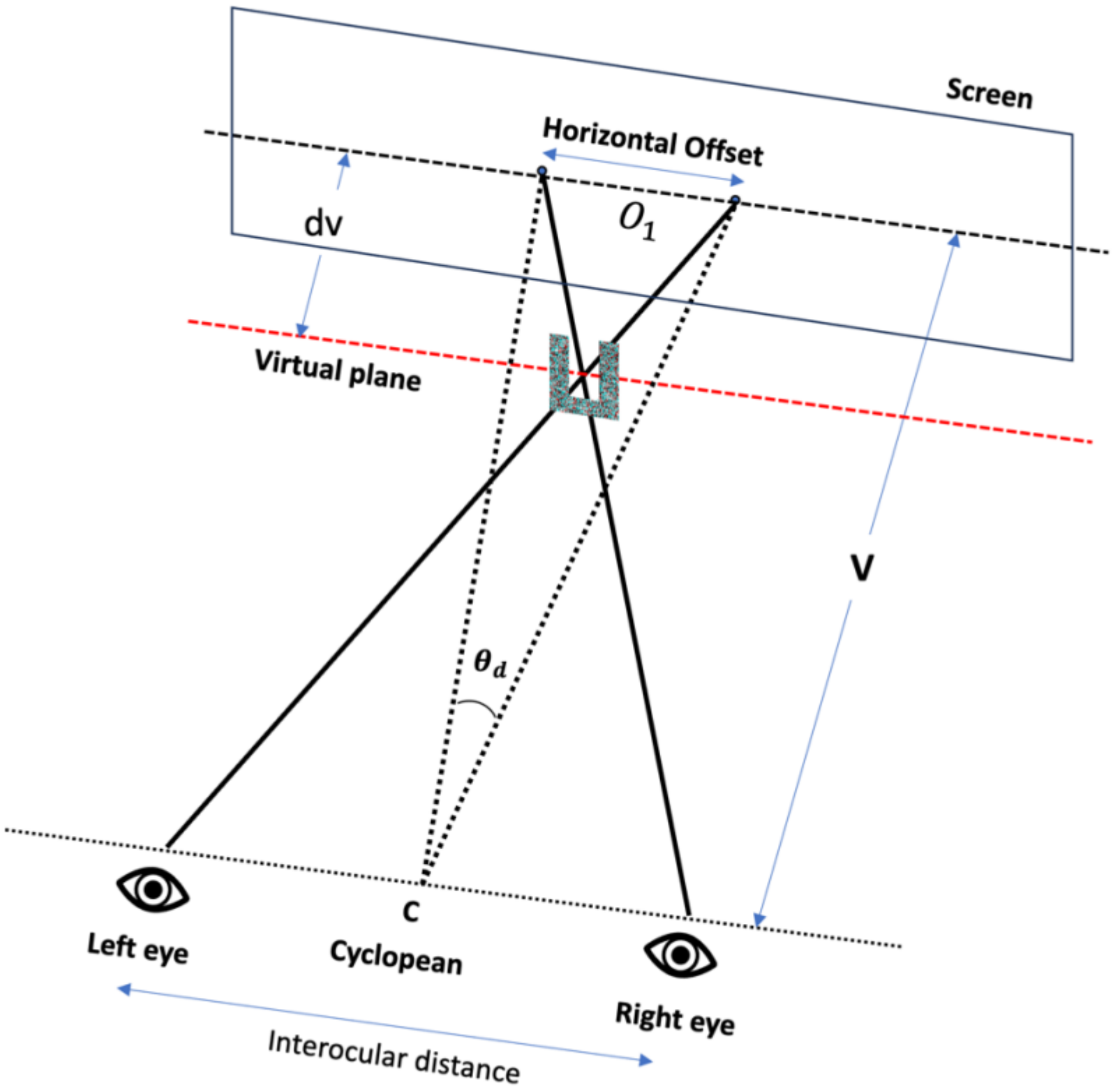

Figure 3. Geometry of the stereopsis pertaining to perceived protruding shape ‘⊔’.

The conversion of effective offset from shift in pixels ($O_1$) to arc-sec ($\theta_{da}$) is done through following steps below:

$$O_1' = \frac{O_1 \times screen\ width\ (mm)}{screen\ pixels\ in\ x\ axis} \tag{13}$$

$$\theta_{dr} = 2 \times \arctan\left(O_1' \times Z\right) \tag{14}$$

$$\theta_{da} = \frac{180}{\pi} \times 3600 \times \theta_{dr} \tag{15}$$

Where Z is the viewer distance, $\theta_{dr}$ is effective offset in radians.

However, this process created unwanted monocular contours which were subsequently removed by horizontally shifting the overlapping textures by an empirically chosen amount ($O_2$ pixels) depending upon $O_1$.

$$O_2 = \begin{cases} -5 \ if \ O_1 \geq 7 \\ -4 \ if \ 6 \leq O_1 < 7 \\ -3 \ if \ 5 \leq O_1 < 6 \\ +3 \ if \ O_1 < 5 \end{cases} \qquad (16)$$

Subsequently, the hidden textures were also shifted by $O_2$ so that the effective offset still remained $O_1$. The luminance of the dots as seen through the red and cyan lenses were 2.4 & 3.3 cd/m$^2$ in dark mode (Ambient illumination: 130 cd/m$^2$) and 7.2 & 9.8 cd/m$^2$ in light mode (Ambient illumination: 20 cd/m$^2$), respectively. The luminance was measured by a Koico Japan photometer (Digital Light Meter Model No. 7402). The default color coding of the textures was red and cyan. However, it can be customized to other chromatically different colors like red and green, red and blue etc, subject to the availability of corresponding 3D anaglyph glasses.

**2.5 Stereoacuity testing procedure**

In our previous work[19], we established that our Digital Stereo Test (DST) is clinically comparable to TNO in a wide range of stereo-impaired populations. However, similar to TNO, DST consisted of 4 fixed level of disparities (480, 240, 120 and 60 arc-sec). In this work, we developed a psychophysical experiment to provide a stereoacuity value on a continuous measurement scale. Our task involved the participants indicating the shape of a hidden texture inside the test scene for a total of 30 trials. Each trial is a scene consisting of RDS stimuli and 4 click buttons of shapes – '⊔', '⊓', '⊏' and '⊐' – at the bottom left of the display rendered by the computing system. Figure 4 illustrates a single trial RDS generated for $O_1$ of 4.59 pixels (763.32 arc-sec when viewed at a distance of 40 cm).

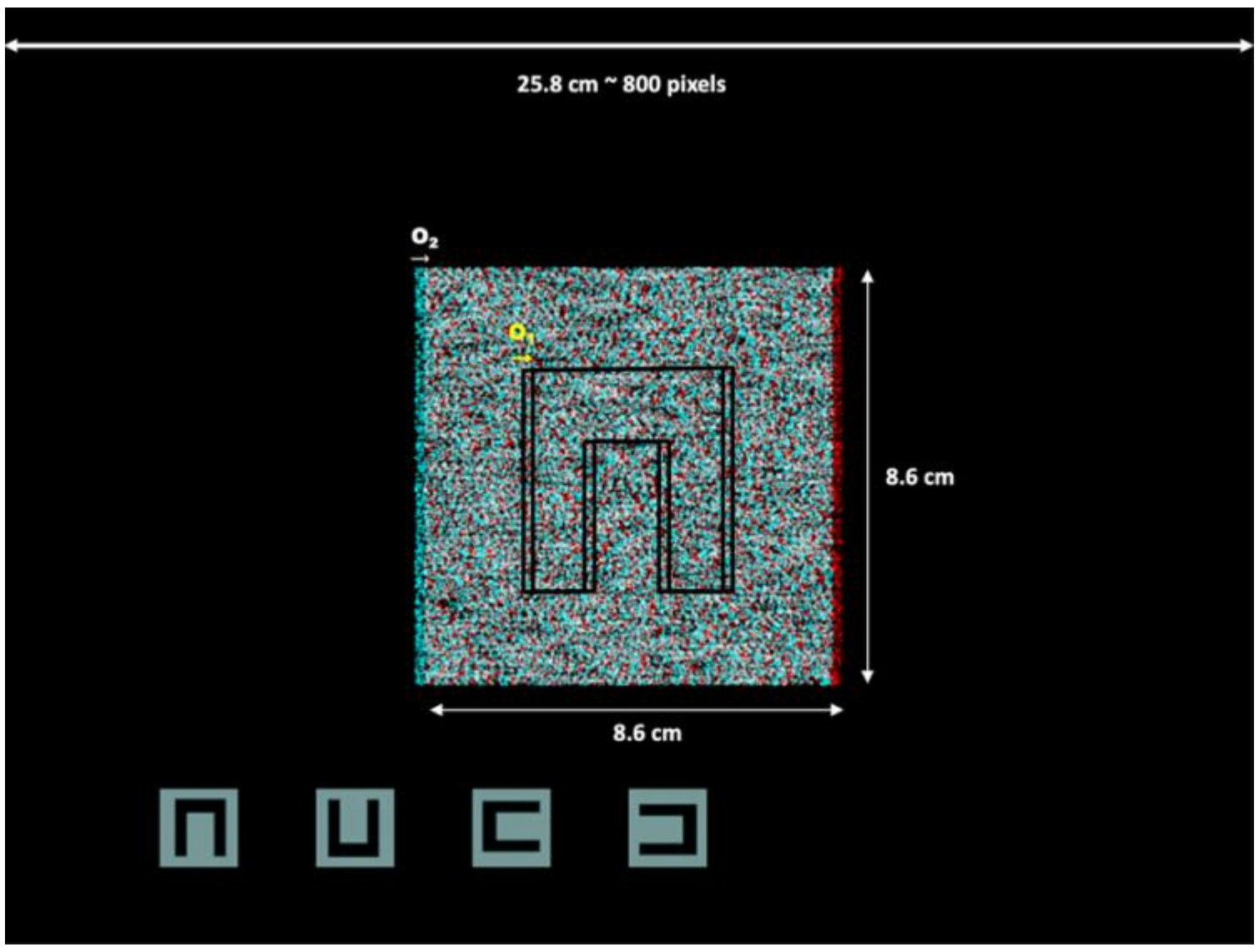

Figure 4. An illustration of a single trial RDS generated for an $O_1$ of 4.59 pixels amounting to 763.32 arc-sec when viewed at a distance of 40 cm. Note: This image is only a screenshot, and the resolution is maintained at $800 \times 600$ pixels in the actual test.

In terms of visualization, each trial differed from every other trial in the amount of $O_1$ towards the projection of a protruding depth effect. We adopted the Expected Entropy Minimization [28] for the computation of $O_1$ for each trial. Below we describe the steps to calculate stereoacuity threshold in detail.

The first step was to define a psychometric function[30] – a function that relates behavioral responses (e.g., proportion correct on a detection task) to a sensory stimulus (e.g., stereoacuity) – generically described in Equation 17.

$$\Psi(x;\ \alpha, \beta, \gamma, \lambda) = \lambda + (1 - \gamma - \lambda)F(x;\ \alpha, \beta) \qquad (17)$$

Where:

$x$: stimulus intensities – [0.1: 0.01: 10]

γ: guess rate (the probability of a correct response when the stimulus is not detected by the underlying stimulus) – 0.25 for 4AFC

λ: lapse rate (the rate at which the participants have the tendency to make an error) – [0: 0.01: 0.04]

$F(x;\ \alpha, \beta)$: the probability of detection by an underlying sensory stimulus. For this purpose, we chose the Weibull distribution as it provides an excellent fit to the true psychometric function[31]. The Weibull cumulative distribution function is given by Equation 18:

$$F(x;\ \alpha, \beta) = 1 - exp(-(\frac{x_i}{\alpha})^{\beta}) \quad (18)$$

Where:

$\alpha$: Threshold – [0.1: 0.01: 10]

$\beta$: Slope (rate of change of performance as a function of stimulus intensity) – 2 to 14

A psychophysical experiment can be seen as a sequence of Bernoulli trials. Psychophysical experiments often use a small range $\{x_1, x_2, \dots, x_k\}$ of distinct stimulus intensities. For a given trial $i$, at a stimulus intensity $x_i$, the probability of getting a correct and incorrect response is given by Equation 19 and 20 respectively:

$$p_i = \Psi(x_i|\alpha, \beta) \quad (19)$$

$$q_i = 1 - \Psi(x_i|\alpha, \beta) \quad (20)$$

Assuming that all trials are independent, then the probabilities of observing an entire sequence of subject responses for correct trials and incorrect trials are given by Equations 21 and 22 respectively:

$$P_i = \prod \Psi(x_i|\alpha,\beta) \qquad (21)$$

$$Q_i = \prod(1 - \Psi(x_i|\alpha,\beta)) \qquad (22)$$

We can aggregate the trials for identical stimulus intensities and define a likelihood function that describes the probability of obtaining correct or incorrect responses for a given set of stimulus intensities, as described in Equation 23.

$$L(r|x,\alpha,\beta) = \prod(\Psi(x_i|\alpha,\beta))^{r_i}\ (1 - \Psi(x_i|\alpha,\beta))^{(1-r_i)} \qquad (23)$$

According to Bayes' rule, the posterior probability can be obtained as:

$$P(\alpha,\beta|x,r) = \frac{P(\alpha,\beta)*ln\ L(r|x,\alpha,\beta)}{\sum_\alpha \sum_\beta P(\alpha,\beta)*ln\ L(r|x,\alpha,\beta)} \qquad (24)$$

wherein *r* is the participant's response and $P(\alpha,\beta)$ is the prior probability density. For the first trial, we define $P(\alpha,\beta)$ as a constant prior. For further trials, the prior is updated as the posterior distribution of the previous trial. After substituting the value of likelihood from Equation 23 in Equation 24, we get:

$$P(\alpha,\beta|x,r) = \frac{P(\alpha,\beta)\{r_i\ ln\ \Psi(x_i)+(1-r_i)\ ln\ (1-\Psi(x_i))\}}{\sum_\alpha \sum_\beta P(\alpha,\beta)\{r_i\ ln\ \Psi(x_i)+(1-r_i)\ ln\ (1-\Psi(x_i))\}} \qquad (25)$$

Next, we compute entropy (a measure of uncertainty associated with $\alpha,\beta$ for all possible combinations of participant responses and stimulus intensities $x_i$ and select the smallest expected entropy for the presentation of the next trial. The entropy of the posterior distribution is defined as:

$$H(r,x) = \sum_{i=1}^{k} P(\alpha,\beta|x_i,r)\ log\ [P(\alpha,\beta|x_i,r)] \qquad (26)$$

Subsequently, the expected value of entropy for stimulus intensity $x_i$ and outcome $r$ is given by:

$$E[H] = \sum_{i=1}^{k} P(\alpha, \beta | x_i, r)\, H(r, x_i) \qquad (27)$$

Finally, the value of $x$ that minimizes $E[H]$ is chosen as the $O_1$ for the next trial. This is repeated for 30 trials and the $O_1$ of the final trial with a correct response is chosen as the stereoacuity threshold of the participant. Assuming that the participant takes at most 10-14 seconds per trial, the total time for computing the stereoacuity threshold typically ranges from 5 to 7 minutes. This duration may vary and could potentially be shorter (depending upon the difficulty of the trial).

**2.6 Statistical Analysis**

For the validity analyses, we computed Spearman correlation coefficients between stereoacuity thresholds obtained in our proposed framework and TNO. We denoted the correlation <0.3 to be weak, between 0.3 and 0.59 to be moderate, and >0.6 to be high[32]. Next, to compute the reliability of the proposed framework with the TNO, we computed the Intra Class Correlation (ICC) coefficient of the stereoacuity thresholds obtained in dark and light modes among all participants (N=97) who performed both 2-step paradigm and the TNO). We did not distinguish between visually healthy controls and participants with real or induced deficits because the reliability of the proposed framework is not affected by the target population. The ICC estimates of less than 0.5 were indicative of poor reliability, values between 0.5 and 0.75 indicated moderate reliability, values between 0.75 and 0.9 indicated good reliability, and values greater than 0.90 indicated excellent reliability[33].

Further, to test the performance of our framework in terms of agreement and homogeneity, we compared the stereoacuity thresholds across dark and light modes using the Bland-Altman method[34] in both the 2-step and 1-step paradigms separately. Then, to assess the systematic differences in the Bland-Altman analyses, we computed non-parametric paired t-tests (Friedman test for repeated measures ANOVA[35]) across dark and light modes in both the 2-step and 1-step paradigms. Additionally, we compared the mean difference between the two modes (dark and light) across both paradigms using the Mann-Whitney U test (two-tailed; $p < 0.05$). Lastly, the homogeneity was evaluated through Spearman correlations between the differences and the means of two tests/measurements, wherein a correlation value of zero would ensure that the differences were independent of the magnitude of the means.

## 3. Results

Table 2 shows a significantly positive correlation across the stereoacuity thresholds obtained during dark and light modes of our paradigm when compared with those of the TNO irrespective of the paradigm type (1 or 2-step) among all participants. To compare correlations across the paradigm type, we used Fisher's z-transformation test and observed that correlation coefficient for the 1-step paradigm was significantly higher (Dark mode: z=2.371, p=0.009; Light mode: z=1.823, p=0.034) than that for the 2-step paradigm in participants with real or induced stereo-deficits. However, no such difference was observed in visually healthy controls (Dark mode: z=0.69, p=0.245; Light mode: z=0.681, p=0.248). Further, to evaluate the agreement of the results of our proposed framework with that of the TNO, we report Bland-Altman plot results for the 1 and 2-step paradigms below.

| Population | N | Type | Spearman r and p with TNO | 95% CI for r | Mode |
|---|---|---|---|---|---|
| Controls | 27 | 2-step | r=0.421; p=0.0289 | [0.048 0.69] | Dark |
| | | 2-step | r=0.422; p=0.0283 | [0.050 0.691] | Light |
| | 24 | 1-step | r=0.575; p=0.0033 | [0.223 0.794] | Dark |
| | | 1-step | r=0.574; p=0.0034 | [0.222 0.794] | Light |
| Real or Induced stereo deficit | 70 | 2-step | r=0.893; p<0.0001 | [0.833 0.933] | Dark |
| | | 2-step | r=0.891; p<0.0001 | [0.83 0.931] | Light |
| | 66 | 1-step | r=0.952; p<0.0001 | [0.923 0.971] | Dark |
| | | 1-step | r=0.941; p<0.0001 | [0.905 0.963] | Light |

Table 2. Correlation across the stereoacuity thresholds obtained during dark and light modes with TNO for 1 and 2-step paradigms among all participants.

### 3.1 1-step paradigm – Only ST

Figures 5a and 6a show box plots summarizing the stereoacuity thresholds obtained through TNO and our framework in dark and light mode conditions. Figure 5a shows the results for the visually healthy participants, while Figure 6a shows the results for the participants with real and induced stereo deficits. Both figures display results collected without calibration. In visually healthy controls, there were no significant differences (Friedman $X_r^2$= 0.167; p=0.683) in stereoacuity thresholds obtained in dark and light conditions, even without calibration prior to ST. The Bland-Altman plot also showed clinically good agreement between the dark and light conditions with a mean difference of -1.1 arc-sec (Limits of Agreement: [-9.5, 7.4], Figure 7a). However, in participants with real or induced stereo deficits, there was a significant difference (Friedman $X_r^2$= 28.015; p<0.00001) in the stereoacuity thresholds obtained in dark and light modes. The stereoacuity values obtained in the light mode overestimated those obtained in the dark mode by 30.3 arc-sec as shown in the Bland-Altman plot (Figure 8a). Moreover, we observed homogeneity across the range of stereoacuity thresholds obtained in dark and light modes in both visually healthy controls (Spearman r = 0.0748, p = 0.728) and participants with real or induced stereo deficits (Spearman r=-0.0663, p = 0.596).

### 3.2 2-step paradigm – AGC and ST

Figure 5b shows a box plot illustrating the summary statistics for the stereoacuity thresholds obtained through TNO (median stereoacuity: 60 arc-sec) and our devised framework in dark (median stereoacuity: 23.3 arc-sec) and light mode (median stereoacuity: 23.4 arc-sec) conditions after calibration for visually healthy controls (N=27). Friedman tests for repeated measures showed no significant differences in stereoacuity thresholds measured across dark and light conditions ($X_r^2$= 3; p=0.083). Similarly, Figure 6b shows the box plot for stereoacuity thresholds obtained for participants with real or induced stereo-deficits (N=70; TNO median: 240 arc-sec; dark mode median: 263.35 arc-sec; light mode median: 268.75 arc-sec). Friedman tests for repeated measures showed no significant differences in stereoacuity thresholds measured across dark and light conditions ($X_r^2$= 3.65; p = 0.055). The Bland-Altman plot shows clinically good agreement between the stereoacuity thresholds obtained in the light mode with those obtained in the dark mode in visually healthy controls, with an overestimation of only 0.12 arc-sec (Limits of Agreement: [-1.36, 1.12]) in the former mode. This difference was 1.8 arc-sec (also clinically acceptable) (Limits of Agreement: [-11.0, 7.5]; Figure 8b) in participants with real or induced stereo deficits. As with 1-step paradigm, we observed no significant correlation between differences and the means of the stereoacuity thresholds during both modes in visually healthy controls (Spearman r=-0.0142, p = 0.94394) and a weak correlation among participants with real or induced stereo deficits (Spearman r = 0.23608, p = 0.049), thereby ensuring homogeneity across the range of stereoacuity thresholds.

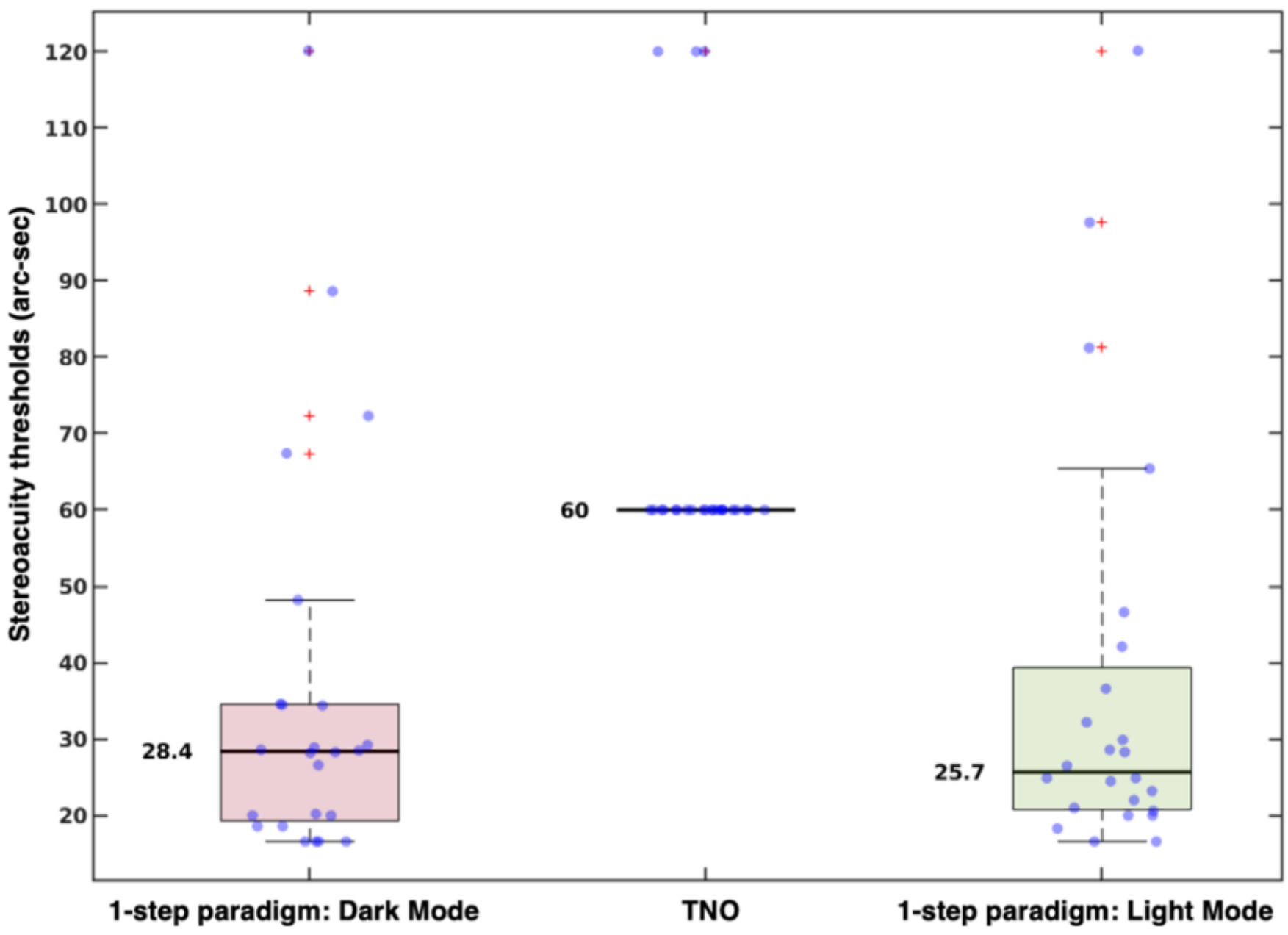

(a)

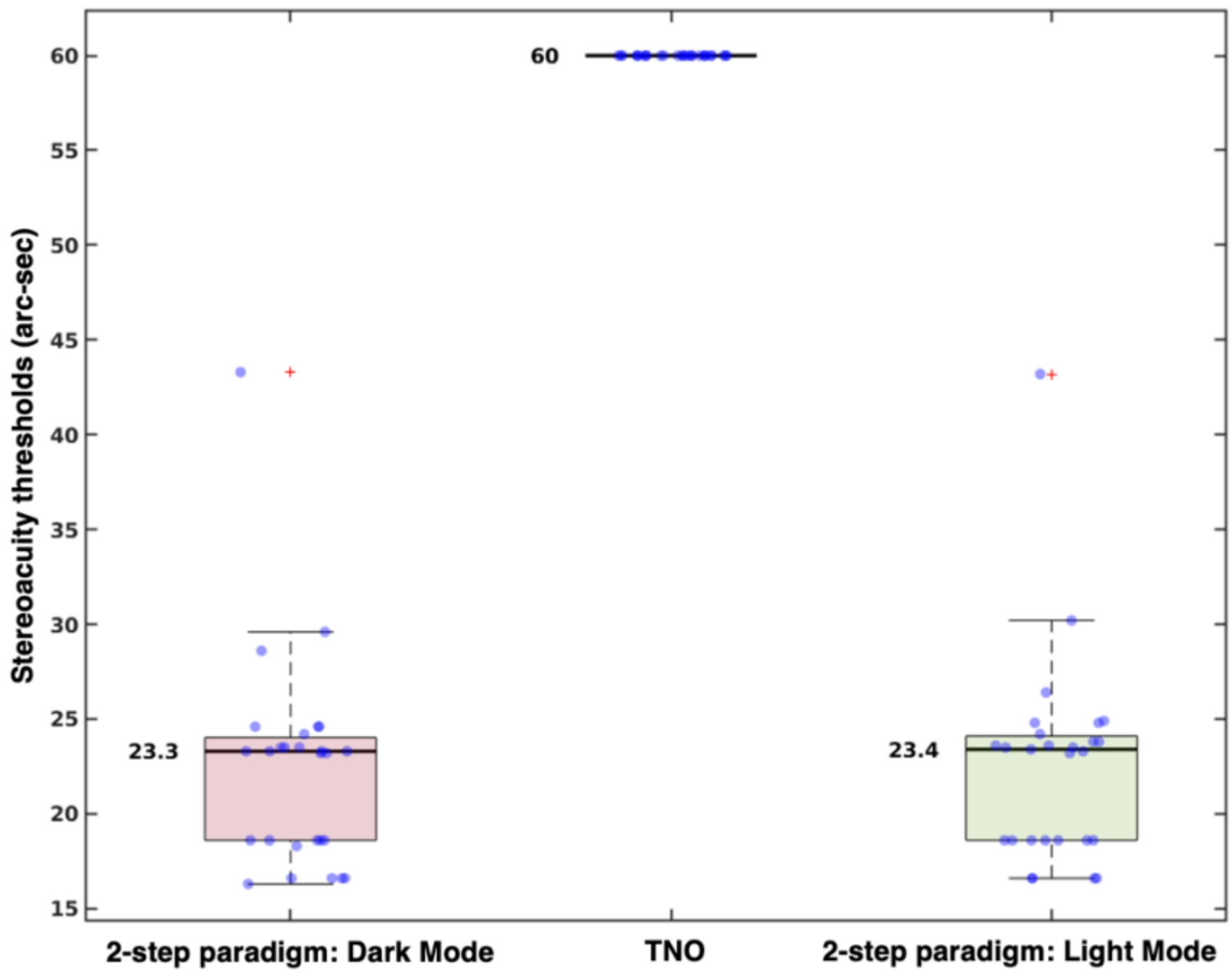

(b)

Figure 5. Box plot illustrating the summary statistics for the stereoacuity thresholds through TNO and our devised framework in dark and light mode conditions before calibration for visually healthy controls (N=24), (b) after calibration for visually healthy controls (N=27).

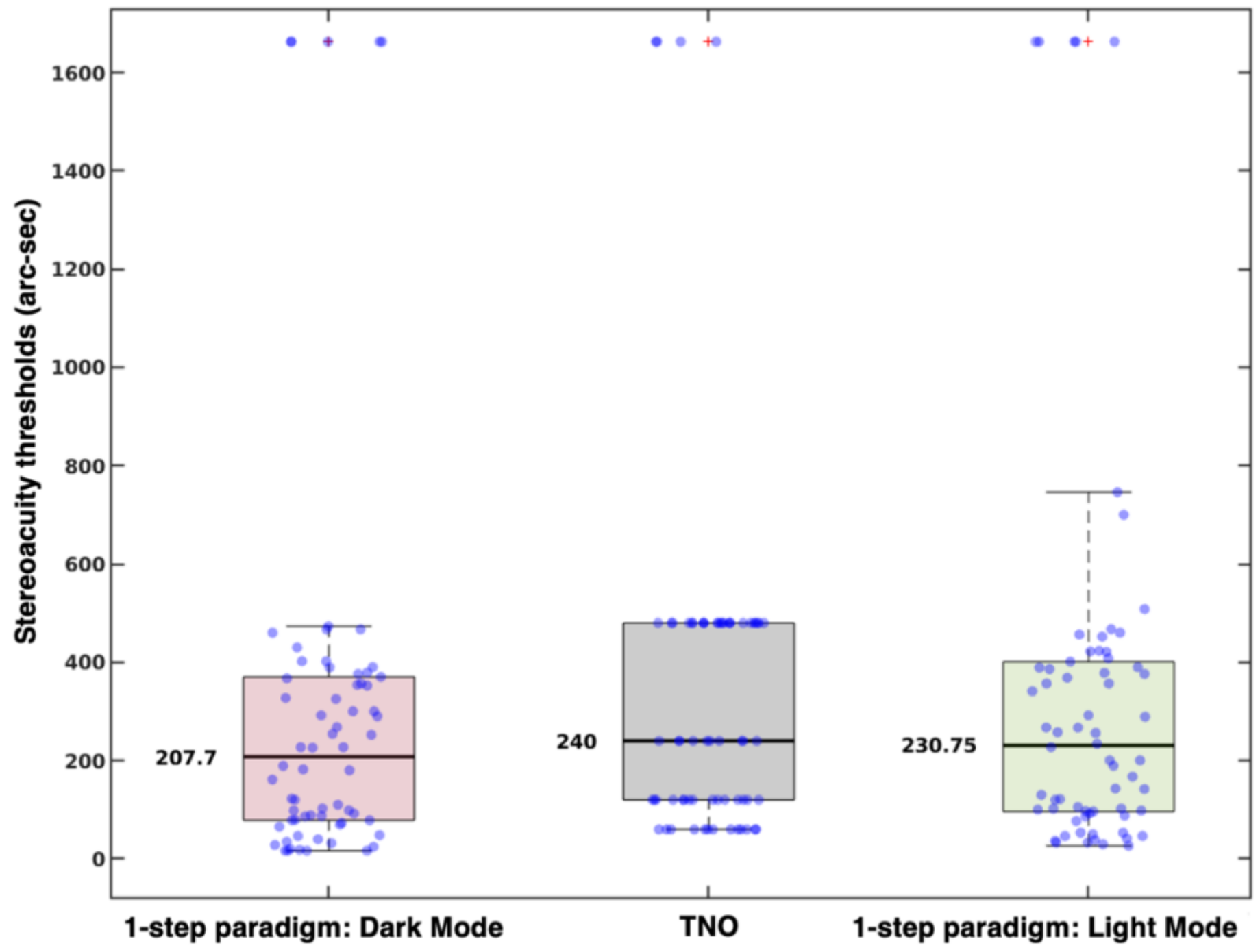

(a)

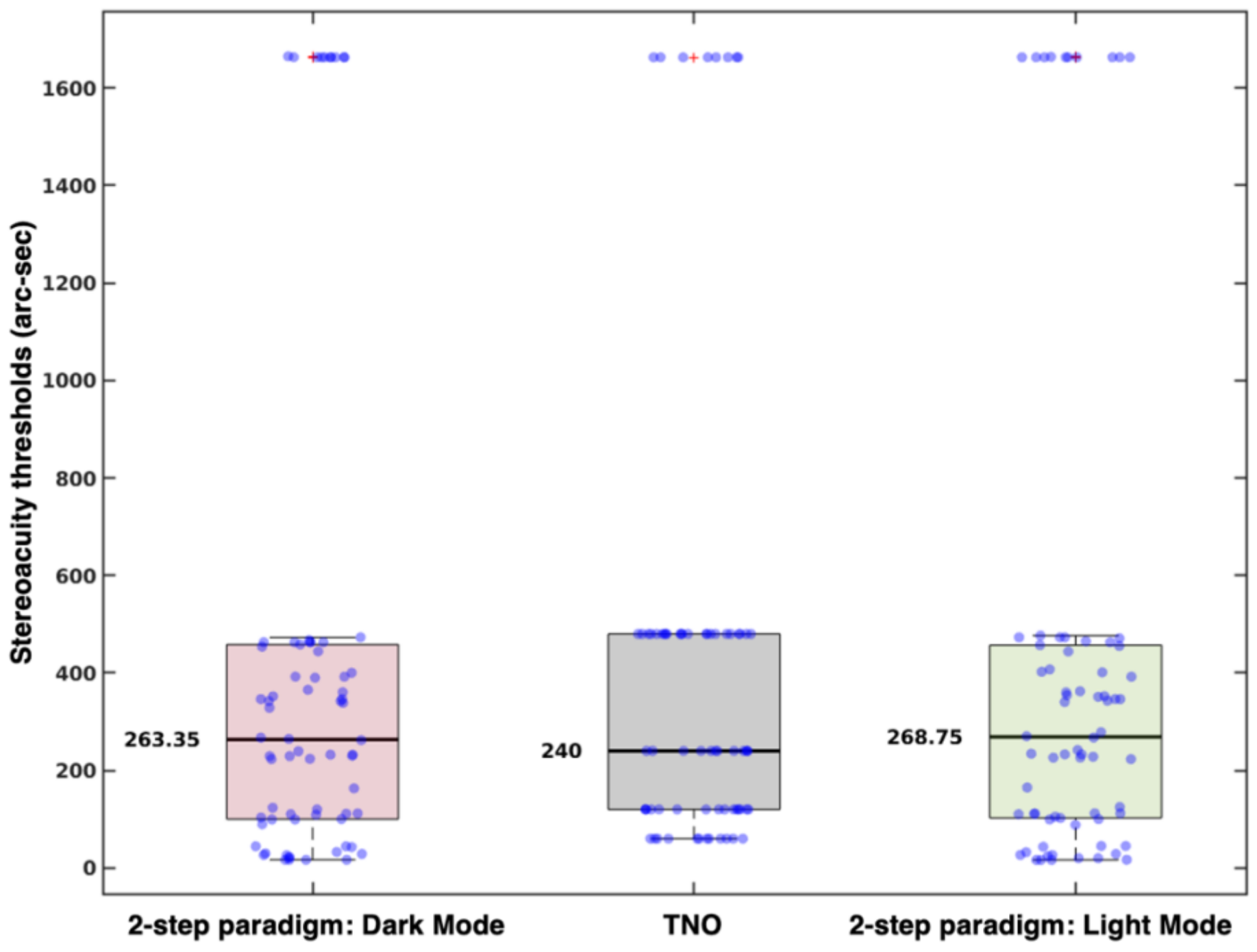

(b)

Figure 6. Box plot illustrating the summary statistics for the stereoacuity thresholds through TNO and our devised framework in dark and light mode conditions before calibration for participants with real or induced stereo-deficits (N=66), (b) after calibration for participants with real or induced stereo-deficits (N=70).

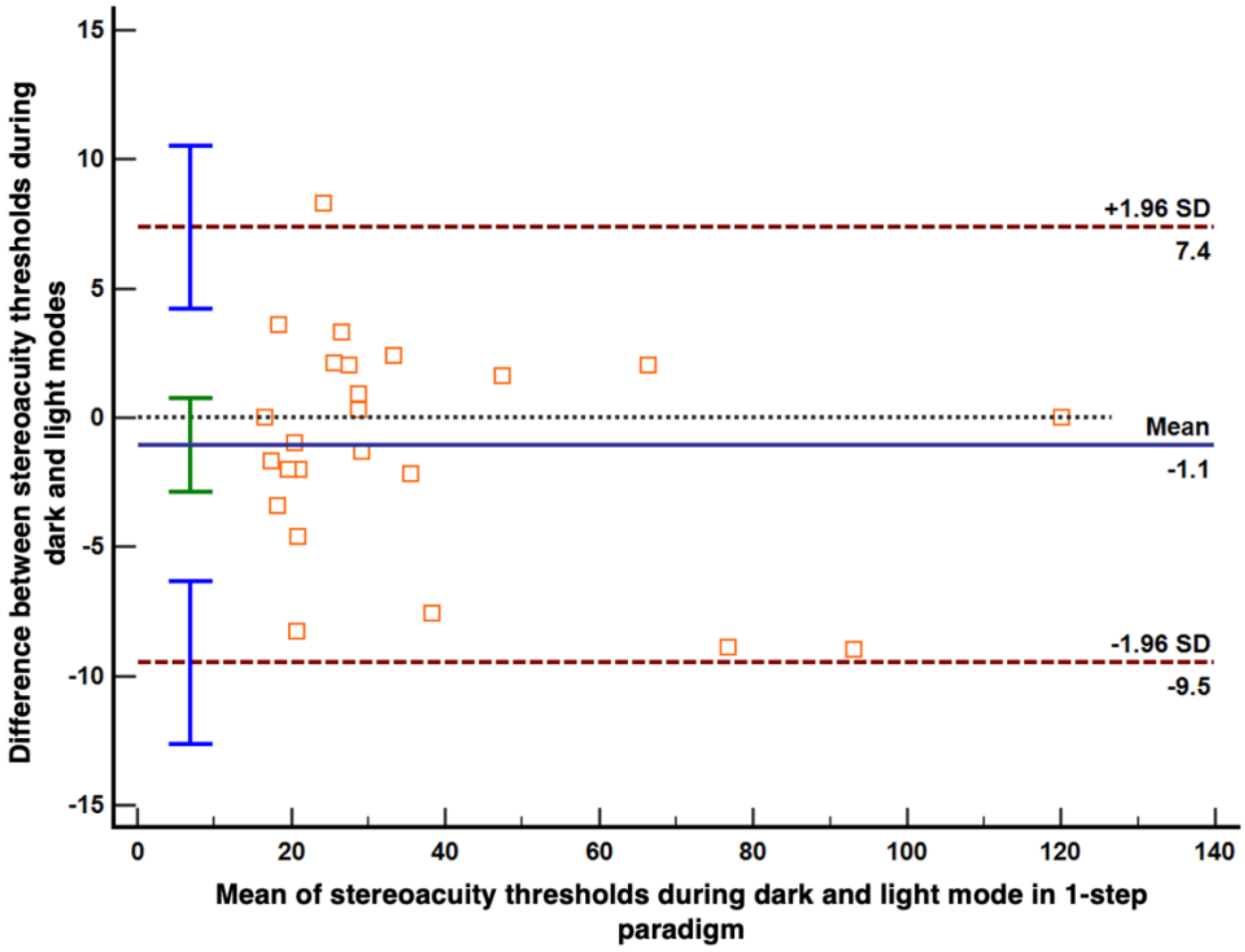

(a)

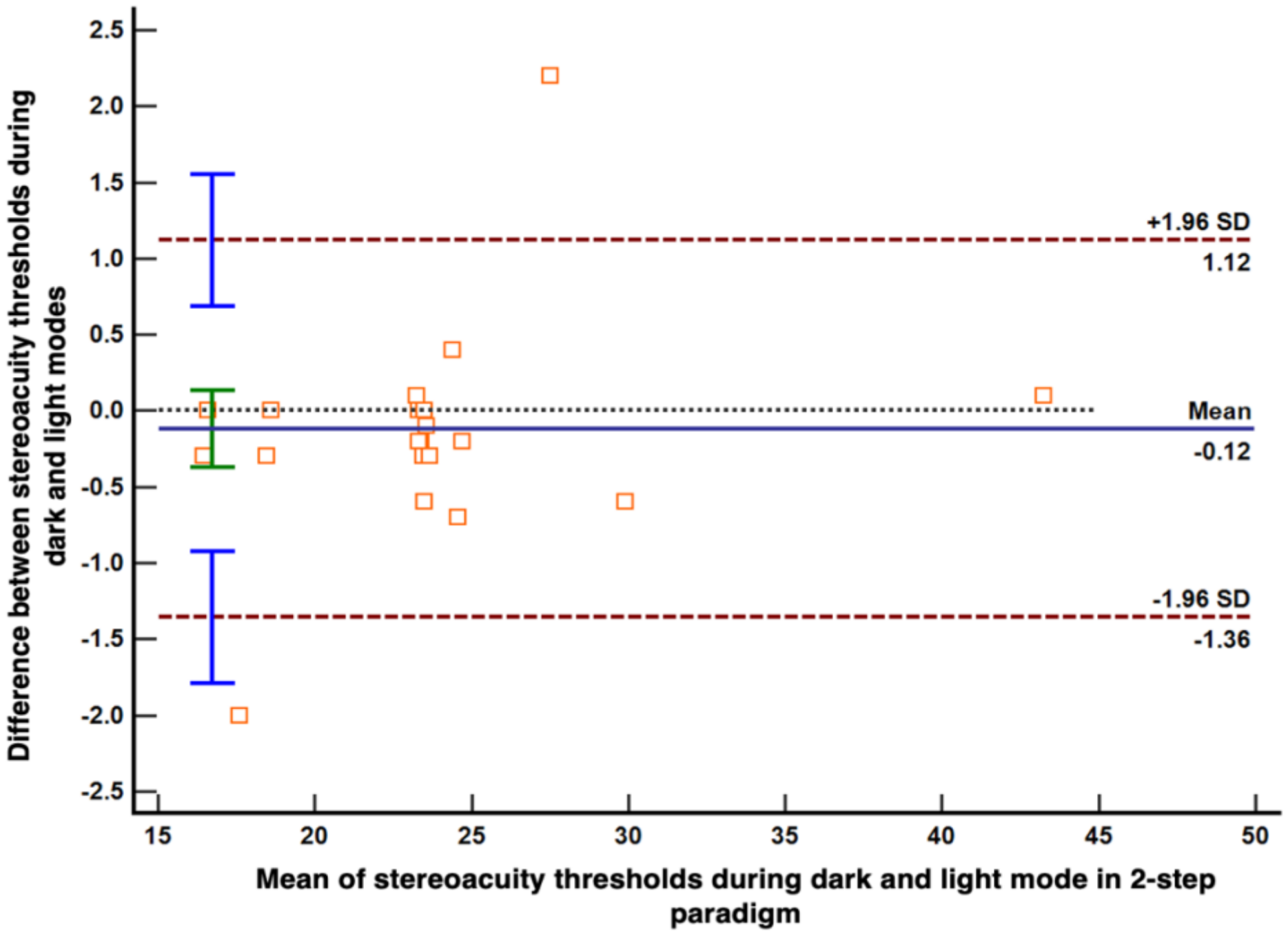

(b)

Figure 7. Bland Altman plot illustrating the comparisons of agreement between the stereoacuity thresholds obtained through TNO and our devised framework in dark and light mode conditions (a) before calibration for visually healthy controls (N=24), (B) after calibration for visually healthy controls (N=27).

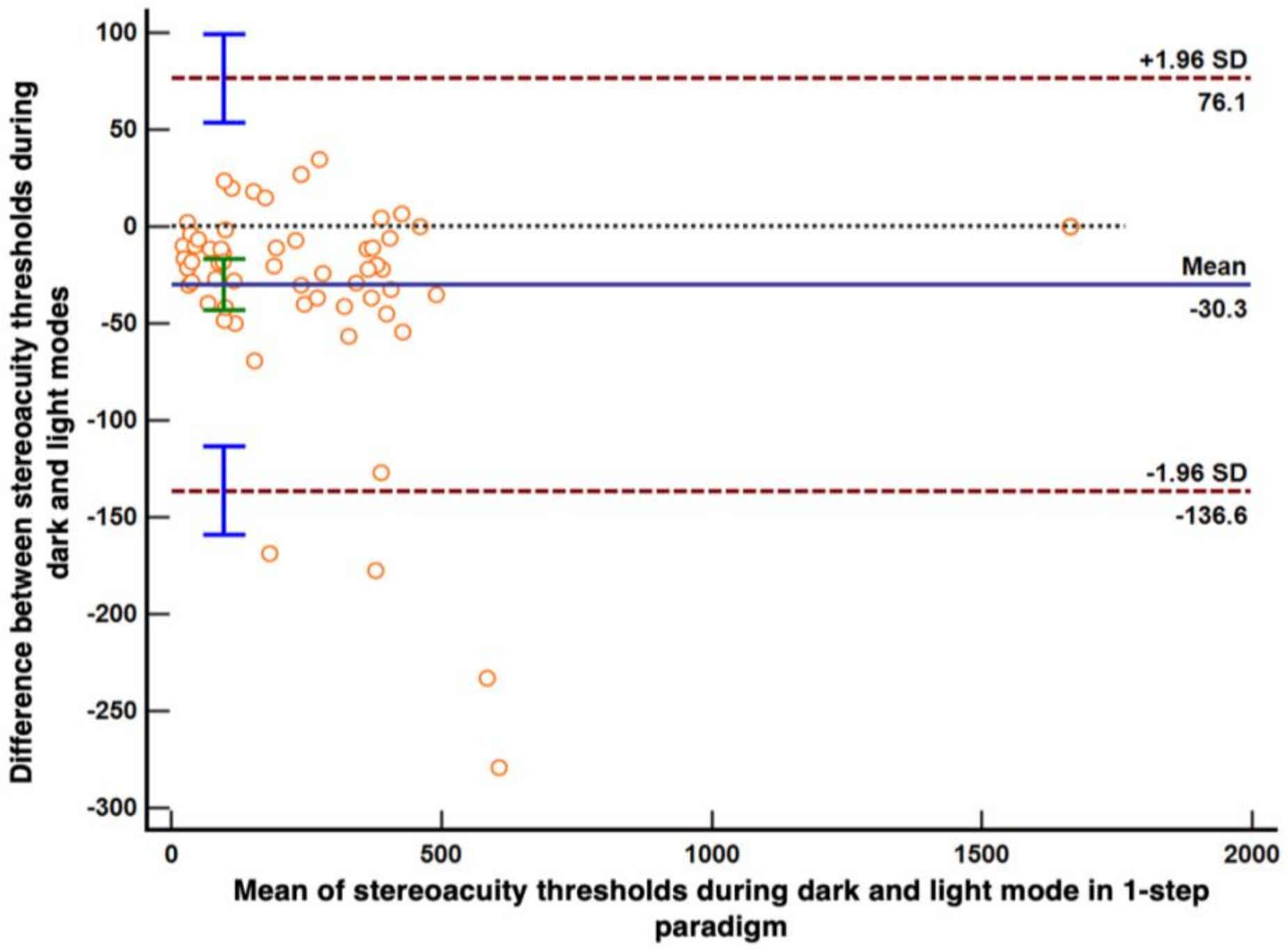

(a)

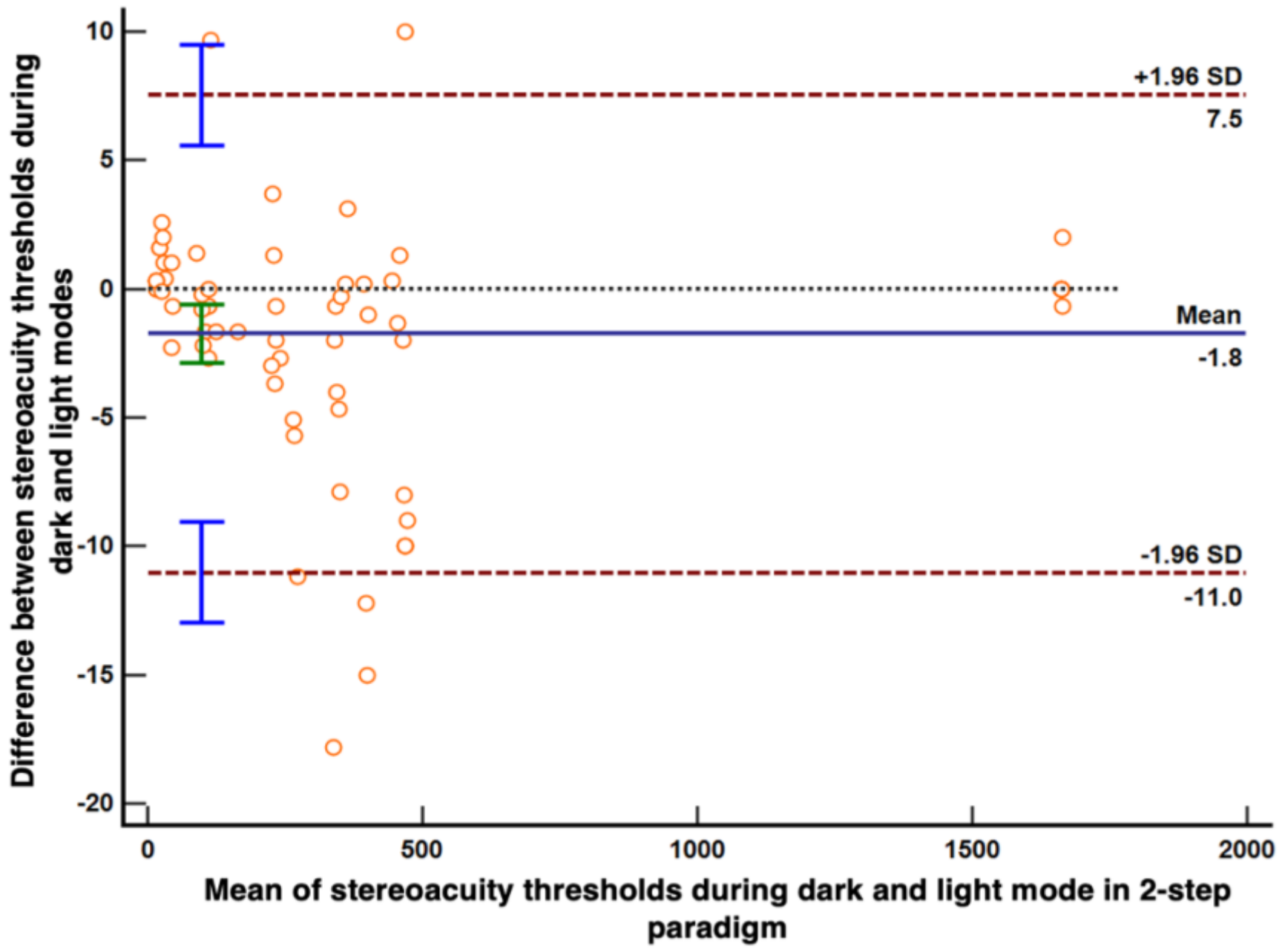

(b)

Figure 8. Bland Altman plot illustrating the comparisons of agreement between the stereoacuity thresholds obtained through TNO and our devised framework in dark and light mode conditions (a) before calibration for participants with real or induced stereo-deficits (N=66) and (b) after calibration for participants with real or induced stereo-deficits (N=70).

### 3.3 Systematic differences in the 2-step vs 1-step paradigm

To understand the effect of lighting conditions on stereoacuity thresholds, we compared the difference in means between the two lighting conditions across both paradigms. For visually healthy participants, we found no significant differences (U=303, z=0.383, p=0.695) in the mean differences of stereoacuity thresholds obtained in light and dark modes for both the 2-step and 1-step paradigms. Contrarily, a significant difference (U=973, z=5.819, p<0.00001) was observed for the real or induced anisometric counterparts.

### 3.4 Reliability of the proposed continuous framework

The ICC estimates were computed for the stereoacuity thresholds obtained from the TNO and those obtained from our framework in the dark and light mode conditions, respectively. Table 3 shows “excellent” (ICC > 0.9) reliability of the proposed framework compared to the TNO. The ICC estimates, and their 95% confidence intervals were calculated based on the mean rating (k=2), absolute agreement in a 2-way mixed effects model using SPSS statistical package version 29.0.2.0 (SPSS Inc, Chicago, IL).

| | | 95% Confidence Interval | | F Test with True Value 0 | | | |
|---|---|---|---|---|---|---|---|
| | ICC | Lower Bound | Upper Bound | Value | df1 | df2 | Sig |
| **Proposed Framework in Light Mode vs TNO** | | | | | | | |
| Average Measure | 0.938 | 0.907 | 0.958 | 15.926 | 96 | 96 | <0.01 |
| **Proposed Framework in Dark Mode vs TNO** | | | | | | | |
| Average Measure | 0.939 | 0.908 | 0.958 | 15.884 | 96 | 96 | <0.01 |

Table 3. ICC estimates based on mean ratings (k=2), Absolute agreement and 2-way mixed effects model for Stereoacuity thresholds obtained during dark and light modes with TNO for our proposed 2-step paradigm among all participants.

## 4. Discussion

The purpose of this study was to provide a comprehensive and continuous assessment of the stereoacuity threshold under ambient lighting. To accomplish this, we used Bayesian inference of psychophysical detection responses combined with a setup of 3D anaglyph glasses. The main findings of our study indicate that the stereoacuity thresholds obtained during the two lighting conditions – dark and light – differed significantly when the AGC was not performed prior to ST, particularly for patients with real or induced stereo-deficits. Further, our framework is able to provide a stereoacuity threshold within 30 trials of the actual test and an additional 15 trials of calibration. The entire process can be completed within 5-10 minutes. Below, we discuss some of the crucial aspects pertaining to the execution of this framework and the interpretation of our results.

### 4.1 Validity and Reliability of the proposed continuous framework

#### 4.1.1 Validity: Comparison with TNO

The thresholds obtained with our proposed continuous framework were significantly positively correlated with TNO under all pairwise conditions. Interestingly, this correlation was higher when calibration was not performed for both controls (Dark mode: r=0.575, p=0.0033; Light mode: r=0.574, p=0.0034 in 1-step paradigm vs Dark mode: r=0.421, p=0.0289; Light mode: r=0.422, p=0.0283 in 2-step paradigm) and participants with real or induced stereo deficits (Dark mode: r=0.952, p<0.0001; Light mode: r=0.941, p<0.0001 in 1-step paradigm vs Dark mode: r=0.893, p<0.0001; Light mode: r=0.891, p<0.0001 in 2-step paradigm). Notably the correlation values tend to be lower in visually healthy controls compared to participants with real or induced stereo deficits. This was because controls were more likely to exhibit 60 arc-sec whereas real or induced stereo deficit participants could exhibit a full range of stereoacuity values ranging from 60 to 480 arc-sec, as measured with TNO. Contrarily the proposed framework was capable of providing a continuous range with finest value measuring at 16.6 arc-sec (compared to 60 arc-sec in TNO). It is also crucial to note that the stereoacuity thresholds obtained in 1-step paradigm were significantly more positively correlated with TNO than those obtained in 2-step paradigm in participants with real or induced stereo-deficits (refer to section 3). This observation indicates that the 1-step paradigm resembles more closely with a fixed level stereo test (i.e., TNO), rendering it less precise than our proposed 2-step paradigm.

#### 4.1.2 Reliability: Comparison with TNO

The proposed framework demonstrated "excellent" reliability in measuring stereo thresholds, irrespective of whether the test was performed in light or dark conditions. The average ICC measure is above 0.9 in both cases; the true values ranged between (95% CI: [0.907, 0.958]) and (95% CI: [0.908, 0.958]) for light and dark conditions respectively (see Table 3). While the reliability is still considered “excellent”[33] in this case, the worst-case ICC estimates (lower bounds) are still above 0.9. These reliability estimates indicate that the thresholds obtained using the proposed framework are comparable to the current gold standard in ophthalmic settings.

### 4.2 Stereoacuity thresholds obtained in dark and light modes during the 1-step paradigm vs the 2-step paradigm

Our analyses reveal a significant finding among participants with real or induced stereo deficits - if calibration is not performed, then the stereoacuity thresholds obtained under dark and light conditions are significantly different. Comparison across both paradigms also show significant differences in the extent to which lighting condition affects stereoacuity thresholds particularly in participants in real or induced stereo deficits. Visually healthy controls did not get affected by the changes in the ambient illumination. This is an interesting observation as numerous studies have suggested that human stereopsis is sufficiently robust against a large range of variations in luminance and contrast[36–39]. These studies however, have been limited to only visually healthy controls[40–42]. Moreover, prior studies were either performed in a completely dark room or did not account for the ambient luminance[26,36]. For instance, Lu Liu et al.[36] noted that contrast did not significantly influence the outcome of a stereoacuity test unless it was extremely low, below 30%. Since typical monitor contrasts do not reach such low levels, the authors suggested that this factor is likely irrelevant to the test results. While they investigated different variations of the display luminance, the ambient luminance was not accounted for.

Besides, the measurement of stereopsis at very high luminance can also lead to poor performance[43]. Rodriguez-Vallejo et al.[18] found that performing stereoacuity test on a digital screen under extreme surrounding illumination would deteriorate its performance. Our framework specifically highlights the need to calibrate display luminance in participants with stereo-deficits to capture subtle changes in stereoacuity. This is crucial because change (or perhaps decrease) in stereoacuity is considered as a significant marker for surgical intervention, highlighting the critical need for its precise detection.

### 4.3 Our proposed continuous framework gives a quick and accurate assessment as compared to previous tests

In our 2-step paradigm, the entire assessment of stereopsis can be done within 5-10 minutes. Previous methods like eRDS v6[15] also provide a continuous framework for ST; however, the total test time is approximately 9 times compared to our framework with an additional requirement of a cumbersome mirror-based stereoscopic setup (45 minutes compared to our 5-10 minutes). In another continuous ST approach, Rodriguez-Vallejo et al.[18] evaluated stereopsis at two testing distances – 0.4m and 3m on an iPad. While their test is capable of measuring distance stereopsis, there are certain limitations: 1) the smallest stereoacuity that can be measured with the limited resolution (2048 × 1536) of iPad was 40 arc-sec as opposed to 16.6 arc-sec in our proposed framework, 2) their testing population is limited to participants with no ocular diseases. Therefore, it is not certain if the device is capable of performing well even in participants with a wide range of stereo-deficits and 3) their test provides more accurate measurements under a fixed room lighting of 945 lux. On the other hand, Portela-Camino et al.[17] recruited a cohort of participants with stereo-deficits in their Computerized Stereoscopic Game test. While their computerized test demonstrated agreement with TNO in stereoacuity thresholds, it was constrained by a limited set of 22 fixed disparity levels. In contrast, our framework overcomes these aforementioned limitations by accounting for ambient lighting, testing with a stereo-deficient population, and incorporating a continuous measurement scale in significantly reduced time.

Although the concept behind the proposed framework can be extended to multiple distances, there are limits on the minimum and maximum disparities that can be presented on our digital display. However, the current disparity range of 16.63 to 1663 arc-sec can be further expanded with an appropriate display size and adjustments to the stimuli and visual angles. Future studies should aim to validate this proof-of-concept using a larger display and test stereoacuity thresholds at various viewing distances.

## 5. Conclusion

We present a new framework that uses 3D anaglyph glasses combined with AGC and Bayesian inference techniques to interactively detect and evaluate stereoacuity thresholds on a continuous measurement scale. We find that the luminance of the test display is indeed a critical factor to detect  subtle changes in stereoacuity specifically in stereo impaired patients. Moreover, we show that the stereoacuity thresholds obtained under the 1-step paradigm align closely with those of the discrete scale of TNO, making it a less precise approach as compared to our proposed 2-step paradigm. Also, the stereoacuity thresholds obtained in our framework are within clinically acceptable ranges of the TNO. Further, stereoacuity thresholds obtained with our proposed framework is able to provide an accurate estimate of stereoacuity within 5-10 minutes. Our framework can be used for early detection and timely management of various eye disorders where stereo vision is impaired.

**Conflict of Interest:** KL, RSS, RS and TKG are listed as inventors on the Indian complete patent application #202411045008 "T.K. Gandhi, K. Lohia, R.S. Soans, R. Saxena (2024), "Method and system for evaluating a stereoacuity threshold of a user in ambient lighting", which is based on the methods described in this manuscript. All other authors declare that the research was conducted in the absence of any commercial or financial relationships that could be considered as a potential conflict of interest.

## Acknowledgment

The authors thank Menka Sharma (AIIMS-Delhi) and Dharam Raj (AIIMS-Delhi) for helping them with the data collection. The funding organizations had no role in the design, conduct or analysis of this research.

## Data Availability

All data used in this study are available from the corresponding author upon reasonable request.

## Credit authorship contribution statement

Kritika Lohia: Conceptualization, Data collection, Methodology, Analysis/Investigation, Visualization, Writing – original draft, review & editing. Rijul Saurabh Soans: Conceptualization, Methodology, Validation, Writing – review & editing. Rohit Saxena: Conceptualization, Funding acquisition, Methodology, Validation, Supervision, Writing – review & editing. Tapan K Gandhi: Conceptualization, Funding acquisition, Methodology, Supervision, Validation, Writing – review & editing.